\documentclass[12pt]{iopart}   
\usepackage{setspace}

\usepackage{iopams}     
\usepackage{graphicx}	
\usepackage{epstopdf}	
\usepackage{url}		
\usepackage[ansinew]{inputenc}
\usepackage[english]{babel}	

\usepackage{textcomp}	                           
\usepackage{booktabs}          
\usepackage[numbers]{natbib}
\usepackage{tipa}

\begin{document}
\title[Treatment of bimodality in proficiency testing of pH in bioethanol matrix]{Treatment of bimodality in proficiency test of pH in bioethanol matrix}

\author{G F Sarmanho$^1$, P P Borges$^1$, I C S Fraga$^1$ and L H C Leal$^2$}

\address{$^1$ National Institute of Metrology, Quality and Technology (Inmetro); Chemical Metrology Division (Dquim); Av. Nossa Senhora das Gra\c cas, 50, Xer\'em, 25250-020, Duque de Caxias, RJ, Brazil.}
\vspace{5pt}
\address{$^2$ National Institute of Metrology, Quality and Technology (Inmetro), Corporate Management Division (Dgcor); R. Santa Alexandrina, 416, Rio Comprido, 20261-232, Rio de Janeiro, RJ, Brazil.}
\vspace{5pt}

\ead{\mailto{gfsarmanho@inmetro.gov.br}}

\begin{abstract}
The pH value in bioethanol is a quality control parameter related to its acidity and to the corrosiveness of vehicle engines when it is used as fuel. In order to verify the comparability and reliability of the measurement of pH in bioethanol matrix among some experienced chemical laboratories, reference material (RM) of bioethanol developed by Inmetro - the Brazilian National Metrology Institute - was used in a proficiency testing (PT) scheme. There was a difference of more than one unit in the value of the pH measured due to the type of internal filling electrolytic solutions (potassium chloride, KCl or lithium chloride, LiCl) from the commercial pH combination electrodes used by the participant laboratories. Therefore, bimodal distribution has occurred from the data of this PT scheme. This work aims to present the possibilities that a PT scheme provider can use to overcome the bimodality problem. Data from the PT of pH in bioethanol were treated by two different statistical approaches: kernel density model and the mixture of distributions. Application of these statistical treatments improved the initial diagnoses of PT provider, by solving bimodality problem and contributing for a better performance evaluation in measuring pH of bioethanol.
\vspace{5pt}

\end{abstract}

{\footnotesize The final publication is available at Springer via \url{http://dx.doi.org/10.1007/s00769-015-1133-4} }


\maketitle

\onehalfspacing                 
\section*{Introduction}
Biofuels contribute to concerns about global warming and they have a significant impact on energy and the environment. Brazil is one of the leading bioethanol producers around the world and its raw material comes from sugarcane, which is a renewable resource. Some of the environmental advantages of using these types of product as fuel in vehicles include reducing the emissions of SO$_2$, CO, hydrocarbons and the amount of greenhouse gas, which contributes to reducing global warming \cite{Goldemberg1999}. In $2013$, world production of bioethanol fuel \cite{RFA} reached about $88\times10^6$ $m^3$. Many world governments are responding to these concerns; the European Union (EU) will require that $10$ \% of all fuels be derived from biomass by $2020$ and the United States of America (USA) the US Energy Independence and Security Act calls for the development of bioethanol and biodiesel  production. specifications regarding  the energy content, purity and origin of biofuels are of great importance \cite{European,EnergyAct}.

The pH of bioethanol is one of the most important parameters of the quality of biofuel \cite{ANP}, since its value establishes the grade of corrosiveness that a motor vehicle can withstand. The pH in bioethanol is also spelled as pHe since it is an operationally defined measurand to differ from pH measured in aqueous solution. In bioethanol, the use of pHe as a measurement of the acid strength is justified since very low levels of strong acids might not always be detected by acidity test and its values are not directly comparable to pH measured on aqueous solutions \cite{ASTMD6423}. However, as the name pH was used in the PT scheme, the pHe measurements will be spelled simply as pH throughout this paper. Its important to emphasize that the values of pH in bioethanol are not directly comparable to pH measured in aqueous solutions \cite{Bates1973,Bates1963}. 

The necessity of developing a certified reference material (CRM) for bioethanol was claimed by the Brazilian producers in order to guarantee its quality and to increase the exportation of their products, since the specified quality parameters for bioethanol \cite{ANP} could be measured with traceability and reliability and accepted by the international trade market.

Because bioethanol is important to the Brazilian economy, since $2006$ Inmetro has focused its attention establishing quality and ensuring accurate and reliable bioethanol measurements by developing and producing RM as well as by providing proficiency testing (PT) for laboratories \cite{PTEthanolInmetro} and CRM for bioethanol \cite{Biorema2010} in order to establish quality and comparable measurements in bioethanol matrix. PT schemes by interlaboratorial comparisons are an important tool used to evaluate the laboratory's ability to competently perform a test or measurement and demonstrate the reliability of the generated results \cite{Albano2014,Thompson2009}.

It is crucial to emphasise that currently three standards \cite{ASTMD6423, ABNT10891, EN15490} are used for measuring pH in ethanol for the international trade. Therefore, different procedures can be carried out in this matrix. There is no harmonization for using a standardised method internationally accepted to perform measurement in ethanol. In addition, as the measurement of pH in ethanol is not comparable to the pH in aqueous solution, the measurement of pH in bioethanol continues to be method dependent.

PT schemes can face problems with deviations of normality assumption, a common supposition used in statistical models for the evaluation of the participant's results. In general, the use of robust statistical approaches to overcome such anomalies can be enough to solve this problem \cite{AMC1989,ISO13528}. However, considering asymmetrical results and/or  the existence of two or more modes -- multimodality -- in the data, such issues can provide incorrect conclusions about participant's performance in inter-laboratory comparison. Thus, using a more realistic consensus value to overcome this problem is recommended \cite{Ellison2009}, being necessary new identification techniques and posterior treatment of the anomalies.

This work aims to present the problem of bimodality that was obtained in a PT scheme related to pH measurements on bioethanol matrix by nineteen Brazilian laboratories, all of which routinely measure that parameter. In addition, in this paper will be discussed the possible statistical treatment that the PT provider could use to overcome the bimodality problem when not expected in PT protocol.

\section*{Experimental}

\subsection*{Samples}
The bioethanol was provided by \textit{Centro de Tecnologia Canavieira - CTC}, an important sugarcane research center established in Piracicaba, S\~ao Paulo. The batch consisted of a $200$ L volume of anhydrous ethanol with a water content determined by Karl Fischer coulometric titration and certified by Inmetro as water mass fraction \textit{w}$(\mathrm{H}_{2}\mathrm{O}) = 0.543 \ \% \pm 0.039 \ \%$ (expanded uncertainty with coverage factor, $k=2$; confidence level of approximately $95$ \%). After homogenization, each sample was placed in a $500$ mL amber borosilicate glass bottle.

\subsection*{Procedure for pH measurements}
Before the measurements, the pH combination electrode assembly was calibrated using CRM pH $4.005$ (Inmetro, Brazil) and a CRM pH $6.865$ (Radiometer \textregistered, trademark from Radiometer Analytical SAS, France). The pH measurements were performed by using a pH meter (Metrohm, Switzerland, model $713$), a pH combination electrode containing KCl $3$ mol L$^{-1}$ as internal solution (Metrohm, Switzerland, model $6.0232.100$), and finally a Pt $100$ temperature sensor (Metrohm, Switzerland, model $6.1103.000$). A magnetic stirrer and a glass jacketed electrochemical cell connected to a thermostatic bath (Marconi, Brazil, $0.13$ $^\circ$C stability) were also used. All measurements were done at $25.0$ $^\circ$C. For each measurement, bioethanol sample was stirred until it reached $25.0$ $^\circ$C. At this point, the stirrer was turned off and after $30$ s the pH value was obtained. At every three measurements, the glass membrane of the pH electrode was rehydrated by immerging it in HCl $1.0$ mol L$^{-1}$ and NaOH $1.0$ mol L$^{-1}$ solutions, alternately.

\subsection*{Reference Material of Bioethanol}
In the studies for developing the RM of bioethanol, the pH was the value of the property chosen. Before using in the PT scheme, homogeneity and stability of the pH in the bioethanol was conducted \cite{Veen2001a,Veen2001b} following the ISO Guide 35 requirements \cite{ISOGUIDE35} as if the RM were a CRM. The combined standard uncertainty for a RM ($u_{\tiny \textnormal{RM}}$), estimated in accordance with GUM \cite{JCGM2008}, was considered as a square root of the sum of the squares of the standard uncertainties related to the homogeneity ($u_{\tiny \textnormal{bb}}$), characterization ($u_{\tiny \textnormal{char}}$) and long-term stability ($u_{\tiny \textnormal{lts}}$) studies. The value of pH in the RM of bioethanol was $6.68 \pm 0.19$ ($k=2$), for a confidence level of approximately $95$ \% at $25.0$ $^\circ$C, and this value was used as the assigned value in the PT scheme.

\subsection*{PT scheme}
Nineteen laboratories participated in measuring the pH in bioethanol \cite{PTEthanolInmetro}. According to the protocol, the participant laboratories were allowed to use their own procedure, i.e. a specific electrode for measuring pH in bioethanol was not defined.

The performance evaluation was pre-defined by the PT provider as recommended in ISO/IEC 17043 \cite{ISO17043}. The z-score was chosen, as expressed by equation (\ref{eq:zscore}):

\begin{equation}\label{eq:zscore}
z_i = \frac{y_i - y_{\tiny \textnormal{pt}}}{ \sigma_{\tiny \textnormal{pt}} } ,
\end{equation}

where, $y_i$ is the result from the laboratory $i$; $y_{\tiny \textnormal{pt}}$ is the assigned value, an estimate of the measurand value; and $\sigma_{\tiny \textnormal{pt}}$ is the standard deviation for proficiency assessment, a dispersion measure used in evaluation of PT results based on the available information.

Among the possible approaches described in ISO 13528 \cite{ISO13528}, it was agreed with the participants (in the PT protocol) that $y_{\tiny \textnormal{pt}} = y_{\tiny \textnormal{RM}}$, where $y_{\tiny \textnormal{RM}}$ is the pH reference assigned by RM from Inmetro, and $\sigma_{\tiny \textnormal{pt}} = s$, where $s$ is the standard deviation of the laboratories.

\section*{Treatment of Multimodality in Proficiency Pesting}
The presence of two or more modes in PT data eventually occurs in analytical chemistry, mainly because different analysis methods are often used. For example, the presence of two measurement methods for measuring the pH of ethanol in this PT scheme; other cases happens when the measurand is operationally defined or even due to contamination of the sample. In all situations, it is recommended \cite{Ellison2009,Yip2009} that PT scheme providers conduct a search for multimodality before using the techniques to evaluate the schemes. The conclusions, and consequently the entire PT scheme, could be invalidated by the existence of multiple modes \cite{Wong2007}.

If no subjective outside information about the problem indicates a natural bias from the laboratory measurements, the search for multimodality can be difficult. The literature mentions some methods that can be used for this purpose as well as methods that can be used for posterior treatment of the multimodality in the PT context: Cofino et al. \cite{Cofino2000,Fearn2004} reported a method based on quantum chemistry for identifying and estimating consensus values without separating data, which was preferable to robust methods; Lowthian and Thompson \cite{Lowthian2002} proposed the use of kernel estimation of densities, a consensus value for each group of observations is estimated, after testing the existence of more than one mode in the data; Cofino et al. \cite{Cofino2005} also proposed an extension of an earlier study \cite{Cofino2000} by including left-censored data in the assessment; Thompson \cite{Thompson2006}, using the Expectation Maximization (EM) algorithm, proposed the use of a mixture of densities model to estimate each of the modes of the PT data. 

Such models are used to identify and estimate the modes and their respective standard errors of PT data. In this context, these measures can be used \cite{Lowthian2002, Thompson2006} as the assigned value $y_{\tiny \textnormal{pt}}$ and the uncertainty of the assigned value $u(y_{\tiny \textnormal{pt}})$, respectively.

Note that although these uncertainties, $u(y_{\tiny \textnormal{pt}})$, are available regarding the dispersion of the assigned value, such measures are not necessarily good estimates of the standard deviation for proficiency testing, $\sigma_{\tiny \textnormal{pt}}$. In general, estimated values for $u (y_{\tiny \textnormal{pt}})$ are very small and might not reflect the actual dispersion expected by participants in the PT scheme. Therefore, the estimated values are not recommended for use in \textit{z}-scores, unless the main objective is to estimate the standard uncertainty of assigned values.

In the present work, two models were applied to pH bimodal PT data: the kernel density model and the mixture of densities model. These models were chosen due to their practicality and simplicity, as demonstrated in other studies. The following sections briefly detail important aspects of both models, emphasizing their practical use in multimodality problems.

\subsection*{Kernel density model}
The kernel density is a non-parametric model \cite{Rosenblatt1956,Parzen1962} used to estimate the probability density function (pdf) from empirical data. Let $y_1,\ldots,y_n$ be an independent and identically distributed (i.i.d.) sample with pdf $f$ (which is desirable to estimate), then the kernel density is given by

\begin{equation}
\hat{f}(y) = \frac{1}{n \cdot h} \sum_{i=i}^{n} \textnormal{K} \left(\frac{y-y_i}{h}\right)  ,
\end{equation}

where the ``kernel'' $\textnormal{K}(\cdot)$ is an univariate function that determines the density shape,  and $h$, the bandwidth, is a smoother parameter that influences the number of modes in the distribution.

Despite the existence of several kinds of kernels (triangular, uniform, etc.), the normal kernel density is the most commonly used due to its mathematical properties. In this case, $\textnormal{K}(\cdot)=$ \textphi$(\cdot)$, where \textphi$(\cdot)$ is the pdf of the standard normal distribution $N(0,1)$; and $h$ can be seen as an additional variance of the data \cite{Lowthian2002} by adding uncertainty to the process of estimating the standard deviation of proficiency testing.

When the kernel is Gaussian, the optimal choice of $h$ is a normal approximation termed ``Silverman's rule of thumb'' \cite{Silverman1998}, which is given by $h \approx 1.06\cdot \hat{\sigma}\cdot n^{-1/5}$, where $\hat{\sigma}$ is the standard deviation of the sample data.

By choosing a kernel and selecting the $h$ value, a smooth curve is estimated $\hat{f}$, and the sample modes  $\mu_j$ will be estimators of the true mode parameters, $j=1,\ldots,m$. The statistical distribution of the sample mode is not well-defined; therefore, the bootstrap technique \cite{Efron1993} is an alternative to estimate the standard error $\textnormal{SE}(\mu_j)$ without the necessity of assumptions regarding the distribution of the estimator.

\subsection*{Mixture of densities}
The mixture of densities model \cite{Aitkin1980} assumes that the data set is a mixture of unknown proportions $p_j$ of results from two or more different populations. Then, the target density is a convex linear combination of a finite number $m$ of known densities $f_j(y)$

\begin{equation}
\hat{f}(y) = \sum_{j=1}^{m} p_jf_j(y)
\end{equation}

where, $0 \leq p_j \leq 1$ and $\sum p_j = 1$. In practical applications, the densities are generally Gaussians $f_j(\cdot) =$ \textphi$(\cdot)$ and the main objective is the estimation of the parameters set ($\mu_j$, $\sigma_j^2$), $j=1,\ldots,m$, of each population and the proportions $p_j$ of each population $j$ in the mixture. 

Under the maximum-likelihood estimation approach, an efficient method to estimate the parameters is the EM algorithnm \cite{Aitkin1980}, an iterative method that circumvents the non-analyticity of the model's likelihood function.

To choose the number of modes ($m$) in the mixture of densities model, a consistent \cite{Kerebin2000} procedure is to compute the Bayesian  Information Criterion (BIC) \cite{Schwarz1978} for each model to be compared. Given the maximized value of likelihood function ($\hat{L}$), the number of parameters to be estimated ($k$) and the size of the sample ($n$), the BIC is calculated by: $\mathit{B} = k\cdot\ln(n) - 2\cdot\ln{\hat{L}}$. The quality of the model fit increases as the BIC is lower.

Finally, it is suggested in \cite{Thompson2006} to use $\mu_j$ as estimates of the assigned value $y$ for each new population $j=1,\ldots,m$. The standard error, used as the uncertainty of the assigned value, is given by $\textnormal{SE}(\mu_j) = \hat{\sigma}_j/\sqrt{n \cdot \hat{p}_j}$.

\section*{Results and Discussion}

\subsection*{pH results from the PT scheme}
Each participant performed $n_i$ measurements of the pH samples received from the PT provider, where $3 \leq n_i \leq 5$, and the mean of these $y_{ij}$ values, $j=1,\ldots,n_i$, were calculated for each laboratory, $i=1,\ldots,n$. Thereby, $y_i = \sum_j y_{ij}$ was the value considered to be the single measurement used to evaluate the PT performance.

All statistical analyses were performed using the \texttt{R} Statistical Software \cite{RCRAN}, a free open-source environment for statistical computing and graphics. The kernel density model estimation was performed by \texttt{density} function of \texttt{base} package, and the mixture of densities model estimation was performed by the \texttt{Mclust} function of the \texttt{mclust} package \cite{mclust,Fraley2001}.

\begin{figure*}[htb]
	\includegraphics[width=5in]{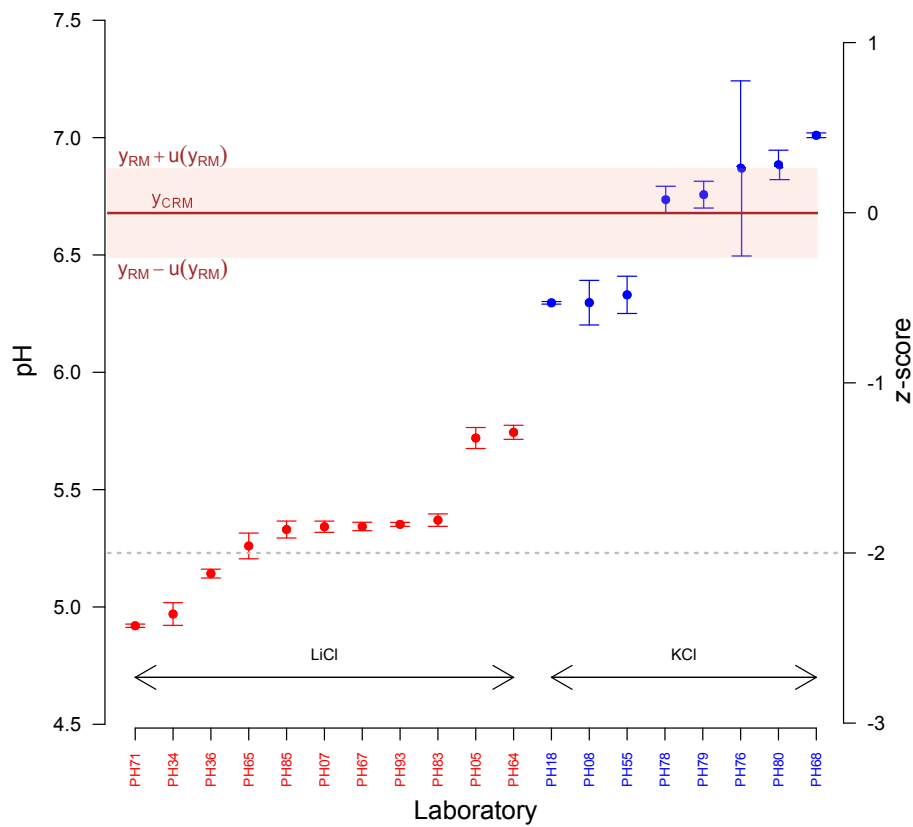}
	\caption[ErrorBarPlot]{Data of pH Proficiency Test: points represents the mean $y_i = \sum_j y_{ij}$ of $j$ measurements informed by the $i$-th laboratory, where $j \leq 5$; the error bars represents the standards deviations of each measurement set, LiCl and KCl respectively. Region around $y_{\tiny \textnormal{CRM}}$ represents the expected range of pH measurements in relation to Reference Material (RM) of bioethanol from Inmetro. In the right y axis is the \textit{z}-score scale, calculated with $Y=6.68$ as assigned value and $s=0.72$ as standard deviation of proficiency testing. Only three participants (PH71, PH34 and PH36) had measurements below the dotted line (\textit{z}-score$=-2$) and so were considered questionable ($2 \leq |z| < 3$).}
	\label{fg:ErrorBarPlot}
\end{figure*}

Figure \ref{fg:ErrorBarPlot} shows the means $y_i$ and the standard deviations (error bars) of each participant, ordered by $y_i$. The differences in the measurement levels between the two groups are observable as shown in red and blue. The laboratories from PH71 to PH64 carried out the pH measurements with electrodes containing saturated LiCl as the internal filling solution, whereas laboratories from PH18 to PH68 used electrodes with a $3.0$ mol L$^{-1}$ KCl internal filling solution. The laboratories that used pH electrodes containing saturated LiCl showed pH measurements that differed by more than 1.0 pH value in comparison to the laboratories that used pH electrodes containing $3.0$ mol L$^{-1}$ KCl.

Comparable results can only be expected for pH electrodes in a non-aqueous medium if standard buffer solutions are available with a composition that is similar to the sample solution. Non-aqueous primary pH standards \cite{Bates1963,Bates1969,Rondinini2002} can be developed by the same procedure used for aqueous solutions \cite{Buck2002}. However, today because of the lack of an ethanol-based reference buffer solution, the pH electrode and pH meter are calibrated in aqueous standard buffer solutions, which leads to large uncertainties due to the medium effect and the phase boundary potential at the interface between the aqueous and the non-aqueous solution \cite{Bates1973}. The traceability of the pH measurement in non-aqueous solutions is far from satisfactory, however, because an international agreement on a primary measurement procedure is required to establish a calibration hierarchy \cite{Spitzer2009}.

As planned in the protocol, the performance evaluation was primarily achieved via \textit{z}-scores. The RM value of pH, $y_{\tiny \textnormal{RM}}$, was the assigned value, $y_{\tiny \textnormal{pt}}$, for the PT scheme ($y_{\tiny \textnormal{pt}} = y_{\tiny \textnormal{RM}} = 6.68$), and the standard deviation of proficiency testing was the standard deviaton of the PT data ($\sigma_{\tiny \textnormal{pt}}=s=0.72$). By using these two values and calculating the scores, the result would not be consistent with the reference of the pH data because of the large standard deviation that comprises all the labs, independent of the internal filling solution used.

Also in figure \ref{fg:ErrorBarPlot}, though $11$ laboratories (those with LiCL as internal filling solution) had mean measurements $y_i$ far from the target $y_{\tiny \textnormal{RM}}$, not one of the total 19 laboratories was considered unsatisfactory ($\vert z \vert > 3$) and only three (PH71, PH76, PH36) would be considered questionable ($2 < \vert z \vert \leq 3$), according to the \textit{z}-score value established in ISO/IEC $17043$ \cite{ISO17043}.

Because $y_{\tiny \textnormal{pt}}$ and $\sigma_{\tiny \textnormal{pt}}$ are unsatisfactory in the sense that both measurements do not reflect a reference and the dispersion for the entire set of data, it is important that the PT provider try to overcome this issue according to the best diagnosis of the data set problem and then choose a new criterion to evaluate the PT scheme.

\subsection*{Searching for multimodality}
To confirm the hypothesis of change in the measured pH value given the electrode internal filling solution type, a previous analysis of multimodality was conducted. 

A Shapiro-Wilk test \cite{Shapiro1965} was performed to confirm the non-normality of the data. The test statistic was $0.88$, with \textit{p}-value$=0.02$, rejecting the normality hypothesis. Grubbs's test  \cite{Grubbs1950} for outlier detection was also performed, but it did not indicate any discrepant value for all $19$ participants.

\begin{figure*}[htb]
	\includegraphics[width=5in]{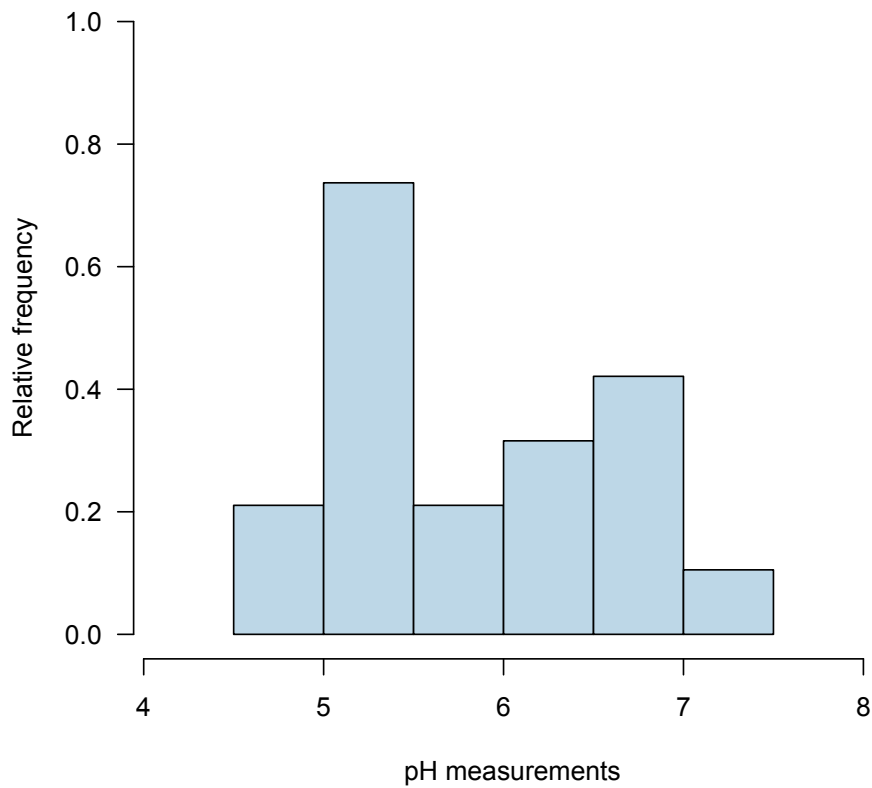}
	\caption[Histogram]{Histogram of PT data. The two peaks of frequency, higher than than others, might be a signal of more than one mode in the data.}
	\label{fg:hist}
\end{figure*}

The histogram of pH PT data is presented in figure \ref{fg:hist}. It suggests two possible groups based on the frequency peaks: one that comes from the laboratories that used pH electrodes with the LiCl filling solution and another from the laboratories that used pH electrodes with the KCl filling solution.

\begin{figure*}[htb]
	\includegraphics[width=5in]{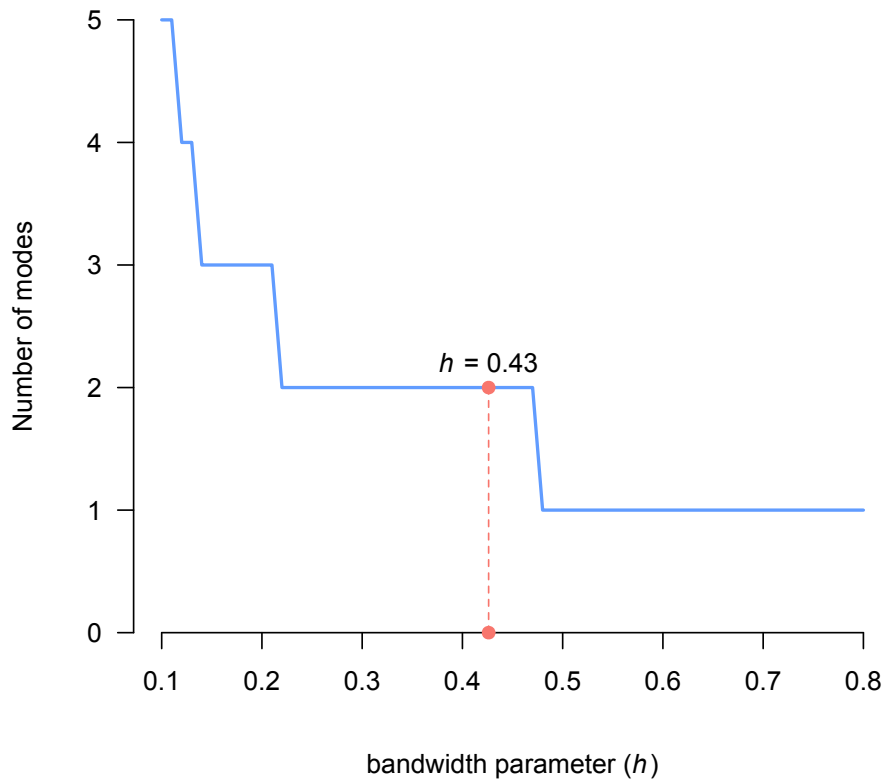}
	\caption[ModesH]{Plot showing the relation between the bandwidth parameter $h$ and the number of modes, according to the PT data. The $h$ value initially estimated by Silverman's rule of thumb was $0.43$, which correspond to $m=2$ modes by the graph.}
	\label{fg:ModesH}
\end{figure*}

Two additional tests were performed to confirm the bimodality in the pH PT data. At first, a Gaussian kernel density model was fitted. The $h$ was estimated by ``Silverman's rule of thumb'' \cite{Silverman1998} as $h \approx 1.06\cdot s\cdot n^{-1/5} = 1.06 \cdot 0.72 \cdot 19^{-1/5} = 0.4261$, and the existence of bimodality was visually tested by plotting the number of modes and the bandwidth parameter $h$, as shown in figure \ref{fg:ModesH}. The estimated parameter $h=0.4261$ suggests two modes to the kernel density, as expected.

A last test was conducted to choose the number of modes based on the BIC. After several Gaussian mixture models have been estimated (by using the EM algorithm) with a different number of mixture components (clusters or groups), the lowest BIC was $\mathit{B}=-40.1$ for $m=2$, confirming that the best model was the one with two components (or number of modes) and equal variances, \textit{i.e.}, $\sigma_j^2 = \sigma_M^2$, for $j=1,2$.

\subsection*{Bimodality treatment of the pH results}

\begin{figure*}[htb]
	\includegraphics[width=5in]{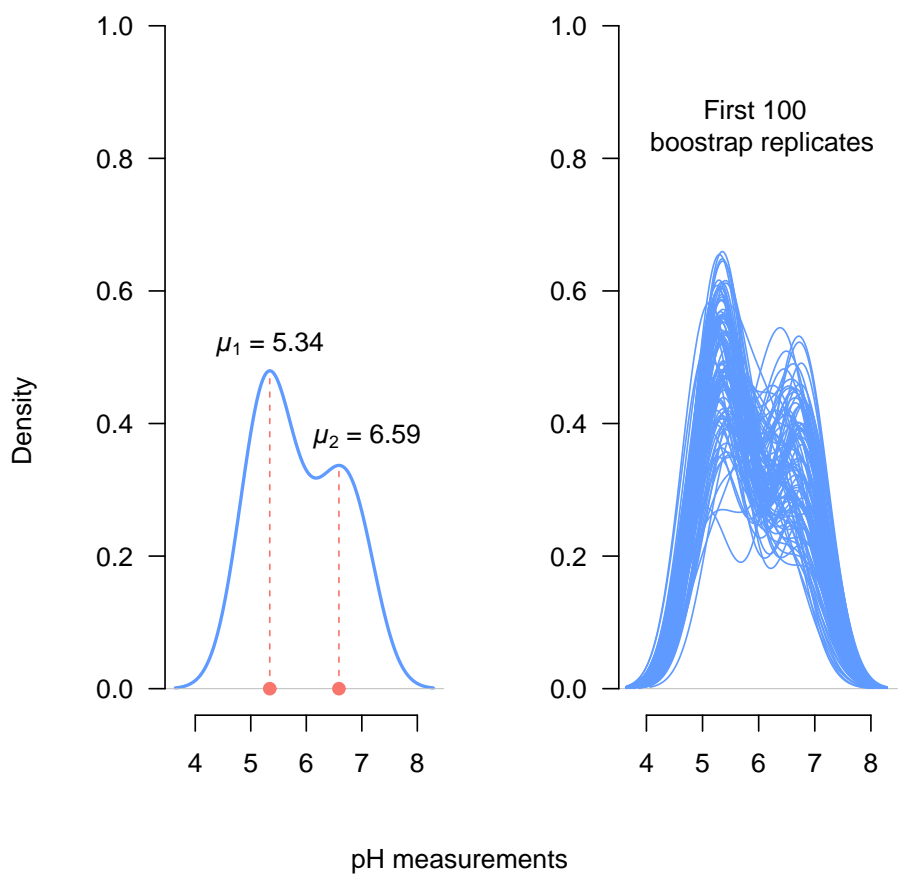}
	\caption[KernelDensity]{(a) Gaussian Kernel Density estimated for pH PT data, with bandwidth parameter $h=0.43$ and modes $\mu_1 = 5.34$ and $\mu_2 = 6.59$ respectively. (b) Superimposed kernel densities of first $100$ bootstrap replicates from pH PT data. It's possible to see most of the curves with two peaks: one between $5$ and $6$, and another between $6$ and $7$.}
	\label{fg:kernel}
\end{figure*}

To overcome the bimodality problem, a Gaussian kernel density with bandwidth parameter $h=0.4261$ was estimated as discussed in the previous section. The first plot (left) in figure \ref{fg:kernel} shows the kernel function estimated with the data as well as the modes $\mu_1=5.34$ and $\mu_2=6.59$. A bootstrap analysis was performed to estimate the standard errors associated with $\mu_1$ and $\mu_2$, which represents the uncertainty of the assigned value. As shown in the second (right) plot of figure \ref{fg:kernel}, the first $100$ simulated curves of $10 000$ are plotted to visualize the behavior of bootstrap samples fitted against the kernel density model for pH PT data. Again, two peaks are visible in the curves, and the graph is interesting because it indicates the variability around the two modes previously estimated by the model. The estimated standards errors were $\textnormal{SE}(\mu_1)=0.07$ and $\textnormal{SE}(\mu_2)=0.10$.

\begin{figure*}[htb]
	\includegraphics[width=5in]{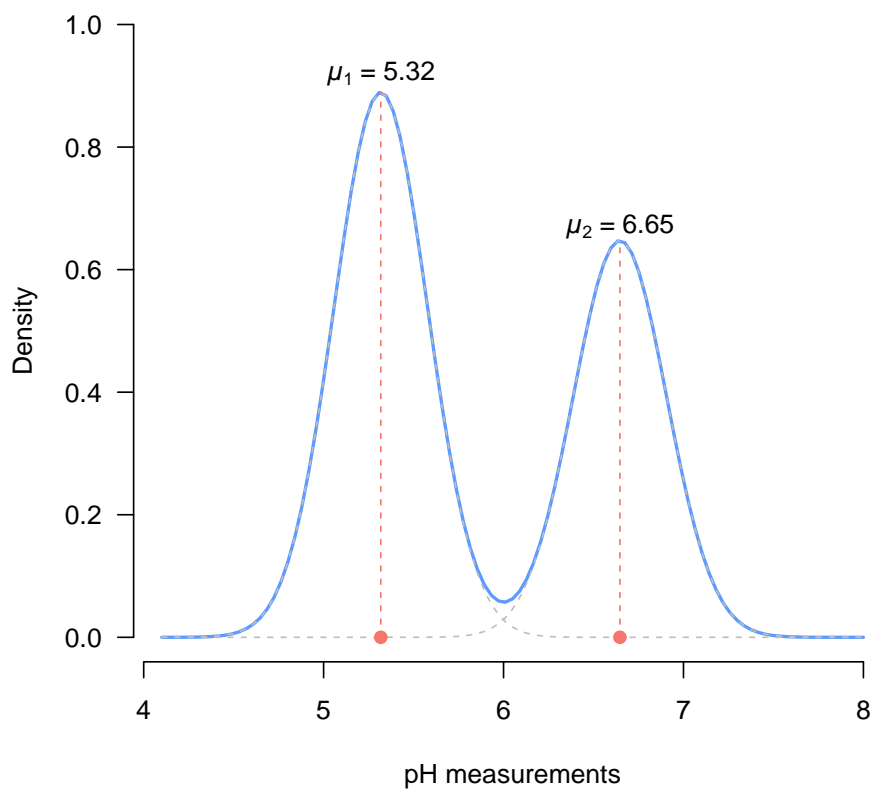}
	\caption[MixtureModel]{Mixture of Gaussians Model estimated for pH PT data, with modes $\mu_1 = 5.32$ and $\mu_2 = 6.65$ , and proportion parameters $p_1=0.58$ and $p_2=0.42$ respectively.}
	\label{fg:mixture}
\end{figure*}

In the same way, the Gaussian mixture model estimation is summarized in figure \ref{fg:mixture}. The mean parameters were $\mu_1=5.34$ and $\mu_2=6.59$; the proportions of the mixtures were $p_1=0.58$ and $p_2=0.42$; and the common variance parameter was $\sigma_M^2=0.26$. Then, the estimated standards errors were $\textnormal{SE}(\mu_1) = 0.08$ and $\textnormal{SE}(\mu_2)=0.09$. The values of the second mode were close to $y_{\tiny \textnormal{RM}}$, which was the initial target of the PT measurements considering KCl as the internal filling solution, but the uncertainties were lower.

Comparing the first graph in figure \ref{fg:kernel} and the graph in Figure \ref{fg:mixture}, the main difference appears in the estimated pdfs for both models. While in the kernel density model a considerable probability mass is evident between the two peaks/modes, in the mixture of densities model, the clear difference between groups is the unique density and also the high probability in both peaks.

After the bimodality diagnosis in the pH results, it is important to establish a new criterion for performance evaluation of the participants because the assigned value previously stated was not consistent with observed data. Because the PT provider had information about the internal filling solution used in the each laboratory analysis, the PT data was split into two groups. Due to the small number of participants in each group (8 and 11), it was suggested to use a robust measure as the estimated standard deviation for proficiency assessment ($\sigma_{\tiny \textnormal{pt}}$) and then perform scores for each group. The normalized median of absolute deviations (herein called MADN) was chosen as robust measure because it is a consistent estimator for the standard deviation and is less influenced by outliers \cite{Huber1981}. It is calculated by $\textnormal{MADN} = 1.483\cdot\textnormal{MAD}$, where $\textnormal{MAD} = \textnormal{median}(\vert y_i-\textnormal{median}(y_i)$ is the median of absolute deviations, a robust measure of the variability of a univariate sample, and $\textnormal{median}(y_i)$ is the median of $y_i$ measurements, $i=1,\ldots,n$. 

\begin{figure*}[htb]
	\includegraphics[width=5in]{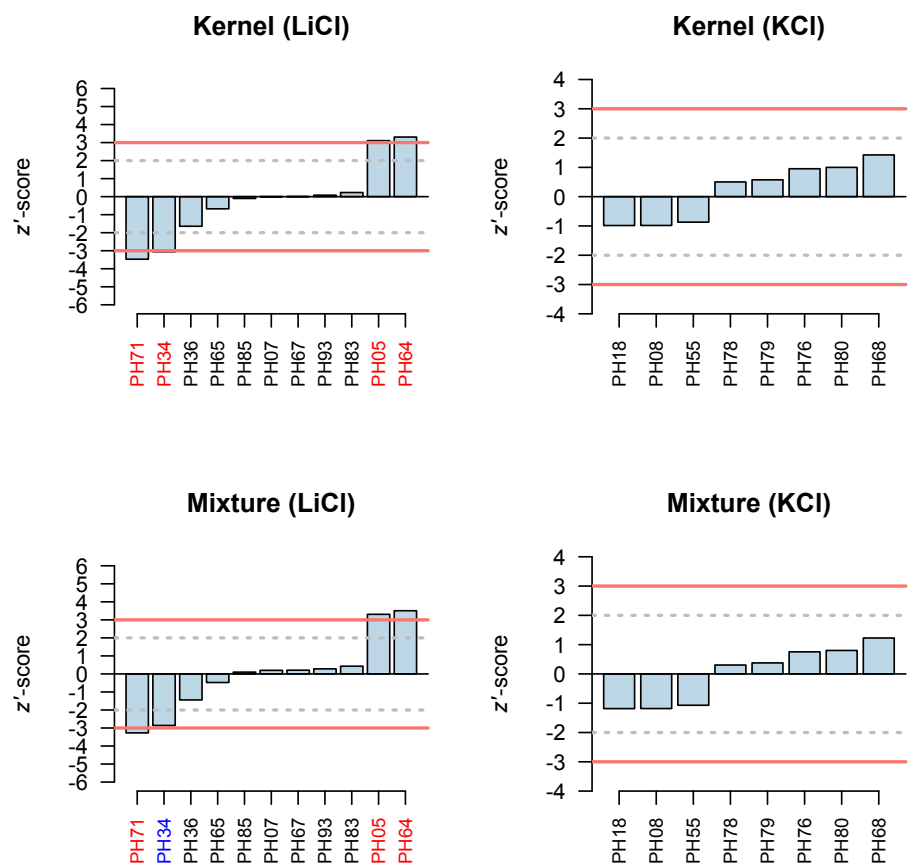}
	\caption[ZscoreFINAL]{\textit{z'}-scores of pH PT data splitted by statistical technique (Kernel density vs mixture models) and by internal filling solution (LiCl vs. KCl)}
	\label{fg:ZscoreFINAL}
\end{figure*}

The calculated $\textnormal{MADN}$ was $0.12$ for LiCL and $0.29$ for KCL. Following recommendations in \cite{ISO13528, LGC2014}, the calculated uncertainties of assigned values were tested to determine whether they were considered negligible by evaluating the expression $u(y_{\tiny \textnormal{pt}}) \leq 0.3 \cdot \sigma_{\tiny \textnormal{pt}}$ in each case. For the four cases (methods versus models), the expression was not met, leading to the conclusion that the uncertainty of the assigned value is not negligible and should be included in the $\sigma_{\tiny \textnormal{pt}}$ for scoring. In this case, the new score is called \textit{z'}-score, simply replacing the denominator of expression \ref{eq:zscore} by $\sqrt{\sigma^2_{\tiny \textnormal{pt}} + u(y_{\tiny \textnormal{pt}})^2}$, but with the same interpretation of the \textit{z}-score. The \textit{z'}-scores for each case (group versus method) are showed in figure \ref{fg:ZscoreFINAL}.  

Note that for both the kernel density model and the mixture model, the outcome of the \textit{z'}-score evaluation was almost identical. For the kernel model, among the 11 laboratories that used LiCl as the internal solution, seven ($63.6$ \%) had satisfactory measurements and the other four participants ($36.4$ \%) had unsatisfactory results. Only one of these four was considered questionable in the mixture of models. This diagnosis can be understood when viewing figure \ref{fg:ErrorBarPlot}, in which the first two and the last two participants had measurements distant from the level of the others measurements.

For the participants who used KCl as the internal solution, none of them all were considered satisfactory. This diagnosis might be explained due to the high variability in the data, also showed in figure \ref{fg:ErrorBarPlot}, and the small number of participants in each case group. In such situations, scores will be given for information only \cite{LGC2014} and should not be used to evaluate the performance of the participants.

\section*{Conclusions}
Results of the PT schemes have shown that the initial approach proposed by the PT provider might not be free of faults. In this case, many interventions had to be made to overcome those problems and better evaluate the participants measurements.

Because the pH of bioethanol results from the PT data showed bimodal distribution based on the type of internal filling solution (LiCl or KCl) used for the pH electrodes, a multimodality approach was applied to compare the two statistical models: (Gaussian) kernel density and a mixture of (Gaussians) distributions. The results from both models were close, which improved the initial diagnoses of the PT provider.

In cases where a previously specified reference value is available (as in the pH PT scheme), that value can be assigned only as a consensus value obtained from the samples belonging to the group that measured the pH using the same method (in this case, the type of internal filling solution), despite the use of the estimated mode that was used in the models in this work. For the other group of samples, the new estimated mode and the respective standard error can be good parameters for the \textit{z}-score evaluation.

Finally, when no information is available about the different measurement methods used by the participants, the same multimodality approach is recommended, that is, applying several models to estimate the modes and the standard errors and then split the participants into groups to perform \textit{z}-score evaluations. The limits for splitting the groups could be the inflection points between two consecutive modes.

This work contributes to overcoming the problem of multimodality that can occur in PT schemes, thereby providing more realistic means for presenting a viable solution for laboratories participants

\nocite{*}					   		
\bibliographystyle{unsrt_spbasic} 	
\bibliography{References}			


\end{document}